\documentclass{emulateapj}
\usepackage{xspace}
\usepackage{amsmath}
\usepackage{hyperref}
\usepackage{framed} 
\usepackage{txfonts}
\def\tsf{\tau_{\rm SF}}
\def\tbw{\tau_{\rm BW}}
\def\euv{\epsilon_{\rm UV}}
\def\Msun{{\rm M}_\odot}
\def\Lsun{{\rm L}_\odot}

\def\ion#1#2{{\rm  #1}\textsc{#2}}
\def\HI{{\ion{H}{I}}}

\def\HeI{{\ion{He}{I}}}
\def\HeII{{\ion{He}{II}}}

\def\H2{{{\rm H}_2}}
\def\lya{Ly$\alpha$}

\def\bm{B20.sf1.uv2.bw10}
\def\bl{B40.sf1.uv1.bw10}

\def\dim#1{\mbox{\,#1}}
\def\hide#1{}

\begin{document}

%=================================================
\title{Cosmic Reionization On Computers I. Design and Calibration of Simulations}
%=================================================

\author{Nickolay Y.\ Gnedin\altaffilmark{1,2,3}}
\altaffiltext{1}{Particle Astrophysics Center, Fermi National Accelerator Laboratory, Batavia, IL 60510, USA; gnedin@fnal.gov}
\altaffiltext{2}{Kavli Institute for Cosmological Physics, The University of Chicago, Chicago, IL 60637 USA;}
\altaffiltext{3}{Department of Astronomy \& Astrophysics, The
  University of Chicago, Chicago, IL 60637 USA} 

\begin{abstract}
Cosmic Reionization On Computers (CROC) is a long-term program of numerical simulations of cosmic reionization. Its goal is to model fully self-consistently (albeit not necessarily from the first principles) all relevant physics, from radiative transfer to gas dynamics and star formation, in simulation volumes of up to 100 comoving Mpc, and with spatial resolution approaching $100\dim{pc}$ in physical units. In this method paper we describe our numerical method, the design of simulations, and the calibration of numerical parameters. Using several sets (ensembles) of simulations in $20h^{-1}\dim{Mpc}$ and $40h^{-1}\dim{Mpc}$ boxes with spatial resolution reaching $125\dim{pc}$ at $z=6$, we are able to match the observed galaxy UV luminosity functions at all redshifts between 6 and 10, as well as obtain reasonable agreement with the observational measurements of the Gunn-Peterson optical depth at $z<6$. 
\end{abstract}

\keywords{cosmology: theory -- cosmology: large-scale structure of universe --
galaxies: formation -- galaxies: intergalactic medium -- methods: numerical}

%----------------------
\section{Introduction}
\label{sec:intro}
%----------------------

Study of cosmic reionization has been highlighted by the last decadal survey as one of the most promising areas of astrophysical research in the current decade. Progress in this area directly influences many other fields of astrophysics, from thermal evolution of the Lyman-$\alpha$ forest to properties of early galaxies.

Because the observational constraints on reionization are limited, theoretical modeling, including numerical simulations, play a relatively larger part in reionization studies than in many other fields of modern astrophysics. Historically, simulations of reionization were mostly confined to two opposite limits: simulations of small spatial volumes with detailed treatment of relevant physics, or large volume simulations with simplified
physical modeling \citep[][and \protect\citet{ng:gt09} for a review of the earlier work]{rei:ipm09,rei:at10,rei:fma11,rei:ais12,rei:sim12}. Both approaches suffer from serious limitations. Small box simulations can model individual ionizing sources with sufficient physical detail, but fail to account for the large-scale correlations between them. Large box simulations include these correlations, but, by ignoring gas dynamics, are not able to model ionizing sources self-consistently. The inability of the simulations to include all relevant scales resulted in a recent surge in semi-numerical and purely analytical approximate methods \citep{reisam:fhz04,reisam:fo05,reisam:cf05,reisam:fmh06,reisam:cf06,reisam:zlm07,reisam:mf07,reisam:aa07,reisam:sv08,reisam:zmm11,reisam:mcf11,reisam:vb11,reisam:mfc11,reisam:kf12,reisam:aa12,reisam:mcf12,reisam:zgl13,reisam:btc13,reisam:rfs13,ng:kg13,reisam:sm14}.

That's where Moore's Law comes to the rescue. The unrelenting exponential increase in the supercomputing power means that sooner or later the gap between small- and large-box simulations is going to be bridged. In fact, \emph{this time is now} - the new generation of supercomputing platforms that have recently been and are planned to be deployed in the US\footnote{For example, ``Stampede'' at Texas Advanced Computing Center, ``Kraken'' at Oak Ridge National Lab, ``Hopper'' and ``Edison'' at Livermore-Berkeley Lab, ``Mira'' at Argonne National Lab, ``Blue Waters'' at NCSA, etc.}, the so-called ``peta-scale'' platforms (since they get close to or exceed $10^{15}$ floating-point-operations per second), are particularly suitable for large-scale simulations of reionization that treat fully self-consistently the radiative transfer of ionizing radiation and gas dynamics.

Taking advantage of this technological progress, we have started a Cosmic Reionization On Computers (CROC) project that aims, over the course of several years, to produce numerical simulations of reionization that model fully self-consistently (albeit not necessarily from the first principles) all relevant physics, from radiative transfer to gas dynamics and star formation, in simulation volumes of up to 100 comoving Mpc and with spatial resolution approaching $100\dim{pc}$ in physical units.

In this first paper in a series, we focus primarily on the technical aspects of our simulations, such as the description of the numerical method, simulation design, and the calibration of simulation parameters. We present the original scientific results from our simulations in the subsequent publications.

%---------------------------
\section{Numerical Tools}
\label{sec:code}
%---------------------------

Our main simulation tool is the Adaptive Refinement Tree (ART) code \citep{misc:k99,misc:kkh02,sims:rzk08}. The ART code is an implementation of the Adaptive Mesh Refinement (AMR) technique with the Fully Threaded Tree data structure \citep{misc:k98}. It includes a wide range of physical processes that make it particularly suitable for modeling cosmic reionization. Specifically, the current version of the code includes the following physical ingredients (in addition to standard ingredients of gravity, dark matter, and gas dynamics).

\emph{Cooling and Heating} of hydrogen and helium is computed ``on the fly'', taking into account all relevant processes in a time-dependent manner, without any assumptions of photoionization or collisional equilibrium. Abundance of heavy elements is tracked self-consistently in ART, but, in a most general case, computing the full dependence of the cooling and heating functions on the incident radiation field is a complex task in itself, and cannot be currently implemented \emph{exactly} in cosmological simulations, unless the radiation field is constant in space \citep{sims:k03,sims:wss09}. 

Since the latter is not a reasonable assumption during reionization or in the ISM of galaxies, ART uses an approximate method of \citet{ng:gh12} that allows to compute the metallicity-dependent part of the cooling and heating functions for an arbitrary time-dependent and spatially-variable radiation field. ART, thus, is able to account for several physical effects that are missed in most other cosmological simulations codes, such as suppression of cooling in strong radiation fields, dependence of the LTE temperature on the radiation spectrum, etc (see \citet{ng:gh12} for some representative examples).

\emph{Radiative Transfer} of ionizing and ultraviolet radiation is currently implemented in ART using the Optically Thin Variable Eddington Tensor (OTVET) approximation of \citet{ng:ga01}. While OTVET is an approximation, it has been 
extensively tested against exact schemes \citep{ng:icam06,ng:icam09}. The Iliev et al.\ tests underscored one undesirable feature of the original ART implementation of the OTVET method - excessive numerical diffusion around ionization fronts. The implementation of the OTVET scheme in ART was substantially revised after those tests, and the currently used approach eliminates numerical diffusion almost completely; a full description of our current implementation of OTVET is presented in Appendix \ref{sec:otvet}. Thus, OTVET remains a highly suitable method for modeling cosmic reionization \citep[see][for detailed discussion of the limitations and inaccuracies of OTVET]{ng:ga01}. 

In our simulations, we include ionizing radiation from stars fully self-consistently (in a time-dependent and spatially-inhomogeneous manner), because it is the primary driver of the reionization process. Other sources of ionizing radiation (quasars, recombination radiation from helium that can ionize hydrogen, bremsstrahlung, etc) we only include in the cosmic background, because these sources are either weakly clustered (helium recombination radiation) or too rare to significantly affect the radiation field in a typical region of the universe (bright quasars). Both components - the radiation from local sources and the radiation from distant sources (i.e.\ cosmic background) - are treated separately in ART, and then combined together to derive a single solution of the radiative transfer equation (see Appendix \ref{sec:otvet}). The advantage of this approach is that it allows to account for the contribution of rare sources (like quasars) to the cosmic background without actually requiring an impractically large simulation volume.

Contributions from helium recombination and bremsstrahlung can be easily computed exactly. Our model for the \emph{quasar contribution} is presented in Appendix \ref{sec:qso}.

Of course, the cosmic background is only important if the mean free path of ionizing photons is sufficiently large, so that the radiation field from distant sources is comparable to or above the radiation field from local sources at a typical location in the universe.

Since we are running several independent realizations for each set of numerical parameters, the post-reionization evolution of the IGM would not be captured correctly if we computed the mean free path for the cosmic background from within one simulation box - periodic boundary conditions will extend that box over the whole universe, whereas it is supposed to represent just one sub-volume of the universe, and only the full set of independent realizations should be treated as a numerical model for the whole universe. Hence, we use the fit from \citet{igm:sc10} to account for Lyman Limit absorptions in the cosmic background; the background is then still subject to local absorptions inside shielded regions, as captured by the radiative transfer solver in Equation (\ref{eq:fu}). Radiation from local ionizing sources is absorbed fully self-consistently with the actually simulated gas distribution in the box (Eq.\ \ref{eq:g}).

\emph{Molecular Hydrogen} chemistry (both gas-phase and dust-based) can be followed in complete detail in ART \citep{ng:gk11}. However, since spatial resolution of our simulations ($\ga100\dim{pc}$) is too coarse to resolve the scale heights of galactic disks, it would make little sense to use the full molecular chemistry module in this work. Instead, we use the fitting formulae of Gnedin \& Draine (2014, in preparation), derived from a large set of small-volume, high resolution simulations, to reliably account for the environmental dependence of the molecular gas on such ISM properties as dust-to-gas ratio or local interstellar radiation field. These fitting formulae are similar to the ones presented in \citet{ng:gk11}, but they also account for the overlap of damping wings of separate absorption lines in the Lyman-Werner band at high molecular column densities.

\emph{Star Formation} cannot yet be modeled from the first principles in cosmological simulations, and needs to be implemented with a phenomenological ``sub-grid" model. In the last several years an important observational advance has been made in understanding star formation on galactic scales. Both, local \citep{ism:lwbb08,sfr:blwb08,sfr:bljo11,sfr:blwb11,sfr:lbbb12,sfr:lwss13} and intermediate redshift \citep{sfr:gtgs10,sfr:dbwd10,sfr:tngc13} observational studies find that the star formation rate surface density on several-hundred-pc scales correlates well, and approximately linear, with the surface density of molecular gas. We use this observed correlation to define our star formation recipe in an entirely empirical manner,
\begin{equation}
  \Sigma_{\rm SFR} = \frac{\Sigma_\H2}{\tsf},
  \label{eq:sfr}
\end{equation}
where $\Sigma_{\rm SFR}$ is the star formation rate surface density, $\Sigma_\H2$ is the surface density of the molecular gas (including the contribution of helium), and $\tsf$ is the molecular gas depletion time scale. We ignore the slightly sub- or super-linear slopes sometimes found in observations, since with our resolution we are only able to resolve a modest range of surface densities where the difference between an exactly linear and a slightly non-linear slopes is negligible.
  
The currently most widely accepted viewpoint is that the depletion time scale $\tsf \approx 1-2\dim{Gyr}$ for normal star-forming galaxies. We use the value of $\tsf = 1.5\dim{Gyr}$ as fiducial, and explore the effect of varying this parameter on our results below.

\emph{Stellar Feedback} is implemented in our simulations with the current industry standard  ``blastwave'' or ``delayed cooling'' model \citep{sims:sdqg09,sims:gbmb10,sims:atm11,sims:bsgk12,sims:aklg13,sims:sbmw13}. While this model is purely phenomenological, it is known to reproduce many of the observed properties of real galaxies well, and, hence, is an appropriate numerical tool at present. The delay time-scale $\tbw$ is a parameter that we vary as a part of the simulation calibration procedure. Our fiducial value of $\tbw=10\dim{Myr}$ is consistent with the usage of this feedback model in the field. 

\emph{Ionizing Radiation from Stars} is the dominant contributor to the global reionization process. The exact amount and spectrum of that radiation depend on stellar IMF and on local absorption inside the galaxy (usually quantified by ``escape fractions''). For our modeled stars we use a fixed Kroupa IMF; the shape of the ionizing spectrum is adopted from Starburts99 modeling \citep{misc:lsgd99} and is plotted in Fig.\ 4 of \citet{ng:rgs02a}. The total UV and ionizing luminosities of a single-age stellar population with mass $m_*$ and metallicity $Z_*$ can be computed with Starburst99; we fit numerical results with the following formula:
\[
  L_{\rm ion} = \euv 1.04\times10^{-4}\frac{m_*c^2 }{Z_*^{0.1}(1+0.27Z_*)} f(t),
\]
where $f(t)$ is such that
\[
  \int_0^t f(t)\,dt = \frac{x\left(0.8+x^2\right)}{1+x\left(0.8+x^2\right)},
\]
and $x=t/(3\dim{Myr})$. At late times ($t\gg10\dim{Myr}$) the ionizing emissivity from a single-age stellar population falls off with time more rapidly than UV light. Our fit behaves in between the ionizing and UV emissivities, since our OTVET implementation requires the same time-dependence of the source function for all radiation bands; that ansatz causes at most a few percent error.

The parameter $\euv$ is unity for the unattenuated stellar output. However, in a numerical simulation with finite spatial resolution some of the absorptions are not accounted for. For example, some of ionizing photons are absorbed in the parent molecular cloud from which stars form, further absorptions occur in the atomic ISM on scales below the effective resolution of the radiative transfer solver. To account for all of these unresolved photon losses, we include the $\euv$ factor and treat it as a free parameters of our model. 

Both, our star formation model and the model for ionizing emissivity ignore the contribution of Pop III stars. This is, necessarily, a simplification. Recent studies of the transition from Pop III to Pop II star formation modes \citep{ng:rgs02a,ng:rgs02b,gals:wa08b,gals:wc09,gals:wtn12,ng:mgg13a,ng:mgg13b} demonstrated, that the transition is rapid and occurs early on, hence the contribution of Pop III stars to the global reionization budget is small. However, if in the future these studies are shown to be incorrect, our simulations should then be considered as a strict lower limit that only accounts for ionizing radiation from the currently observed stars and quasars. 

Another source of ionizing photons that we do not include is annihilation radiation from dark matter. We leave exploring this reionization source to future work.

%---------------------------
\section{Design of the Simulations}
\label{sec:sims}
%---------------------------

\subsection{Resolution Requirements}
\label{sec:table}

The spatial resolution is constrained by the range of scales on which our star formation model is valid, hence we do not vary it in this paper. The mass resolution is a simulation parameter, and, hence, needs to be set carefully. There are several physical arguments that can be used to pick up important mass scales, but the eventual choice of the mass resolution must be done in a convergence study.

The most strict requirement is that all halos that can cool via atomic hydrogen line cooling ($M_{\rm tot}\ga(1-2)\times10^8\Msun$) need to be resolved \citep{ng:go97,ng:g00a,rei:impm06,rei:tc07,rei:zlmd07,rei:mlzd07,rei:tcl08,rei:pss09}. That requirement may be an overkill, though, since dwarf galaxies in the Local Group contain only a few stars \citep[e.g.][and references therein]{dsh:k09} - if these galaxies are typical of dwarf galaxies at high redshift, the mass resolution requirements may be relaxed.

We present the detailed analysis of the mass convergence properties of our simulations in the Appendix \ref{sec:massconv}. The optimal resolution for our simulation, measured as the total ``equivalent particle mass'' $M_1$ (the sum of masses of a single dark matter particle and an average mass of a baryonic cell in the absence of refinement - or the mass of a single dark matter particle compensated by the ratio of $\Omega_{\rm M}/\Omega_{\rm DM}$), would be about $10^6\Msun$, which is equivalent of resolving a, say, $20h^{-1}\dim{Mpc}$ box (in comoving units) with $1024^3$ dark matter particles. As Appendix \ref{sec:massconv} shows, such a simulation would account for 80\% of all ionizing photons. 

Unfortunately, such a resolution is still too computationally expensive at present, so as our fiducial mass resolution we adopt eight times coarser resolution of $7\times10^6\Msun$ (resolving $20h^{-1}\dim{Mpc}$ box with $512^3$ dark matter particles). With this coarser resolution we account for about 55\% of all ionizing photons and for 70\% of ionizing emissivity at $z=6$. At present such precision is sufficient to match the existing observational data; by the time the JWST significantly expands observational samples, faster supercomputing platforms will allow us to increase our mass resoluton to the equivalent of $1024^3$ particles in a $20h^{-1}\dim{Mpc}$ box.

Because in ART the number of dark matter particles throughout the simulation remains fixed, while cells of the adaptive mesh are created and destroyed dynamically, it is more convenient to quantify the resolution in terms of the number of dark matter particles. Each simulation starts with the same number of adaptive mesh cells as dark mater particles, to ensure the consistency of the mass resolution in two main gravitating components. As the simulation proceeds, the number of cells usually grows with time, so that by the end of the simulation the number of cells is a factor of several higher than the number of dark matter particles.

\begin{table}[b]
\caption{Simulation Parameters\label{tab:sim}}
\centering
\begin{tabular}{llll}
\hline\hline\\
Run & Box size & Resolution & Number of \\
   & (comoving) & (proper at $z=6$) & DM particles \\
\\
\hline\\
\underline{Fiducial Series} \\
\\%\hline\\
``B20'' & $20h^{-1}\dim{Mpc}$ & $125\dim{pc}$ & $512^3$ \\
``B40'' & $40h^{-1}\dim{Mpc}$ & $125\dim{pc}$ & $1024^3$ \\
``B80'' & $80h^{-1}\dim{Mpc}$ & $125\dim{pc}$ & $2048^3$ \\
\\
%\hline\\
\underline{Convergence Study} \\
\\
``B20LR'' & $20h^{-1}\dim{Mpc}$ & $125\dim{pc}$ & $256^3$ \\
``B20HR'' & $20h^{-1}\dim{Mpc}$ & $125\dim{pc}$ & $1024^3$ \\
\\
\hline\\
\end{tabular}
\end{table}

Our fiducial simulation series is presented in Table \ref{tab:sim}. Each of the simulations in the series is run with additional 6 levels of refinement (except B20HR, which is run with 5 levels of refinement, to maintain the same spatial resolution as other runs), achieving the same cell size of $125\dim{pc}$ in proper units at $z=6$ ($145\dim{pc}$ at $z=5$), with the actual spatial resolution being a factor $\sim3$ worse. Such resolution is well matched to the range of scales on which the star formation model given by Equation (\ref{eq:sfr}) is observationally tested.

In this paper we only present results from $20h^{-1}\dim{Mpc}$ and $40h^{-1}\dim{Mpc}$ boxes. A simulation with the $80h^{-1}\dim{Mpc}$ is currently feasible to complete on the largest available machines, but it is sufficiently computationally expensive (requiring 20-30 million CPU hours depending on the platform); with the computational resources available to us, we will only be able to afford one per year beginning with 2014.

\subsection{Running Sets of Simulations}
\label{sec:dcmode}

Even our largest ``B80'' simulations, with the $80h^{-1}\dim{Mpc}$ box size, will be only marginally large enough for obtaining convergent results on the distribution of sizes of ionized regions or on observable properties of high redshift galaxies. Thus, in order to extend the reach of our simulations, we run sets of independent realizations of initial conditions for each particular choice of the box size and simulation parameters, accounting for the cosmic variance on the scale of the box size. 

This variance is commonly referred to as the ``DC mode''. The ART code supports the DC mode in arbitrary cosmology without any approximations, following the method described in \citet{ng:gkr11}. In fact, as has been shown in that paper, if a set of independent realizations with a given box size properly accounts for the DC mode, it becomes statistically equivalent to a simulation with a several times larger box. 

\begin{table}[t]
\caption{Completed Simulation Sets\label{tab:sets}}
\centering
\begin{tabular}{lllllc}
\hline\hline\\
Set Id & $\tsf$ & $\euv$ & $\tbw$ & Stopping & Number of \\
 & (Gyr) & & (Myr) & redshift & realizations \\
\\
\hline\\
\underline{$20h^{-1}\dim{Mpc}$ boxes} \\
\\%\hline\\
~~~~B20.sf1.uv1.bw10 & 1.5  & 0.1 & 10 & 5 & 6 [A-F] \\
~~~~{\bf \bm}        & 1.5  & 0.2 & 10 & 5 & 6 [A-F] \\
~~~~B20.sf1.uv4.bw10 & 1.5  & 0.4 & 10 & 5 & 3 [D-F] \\
~~~~B20.sf1.uv2.bw40 & 1.5  & 0.2 & 40 & 5.7 & 3 [D-F] \\
~~~~B20.sf2.uv2.bw10 & 0.75 & 0.2 & 10 & 5.7 & 3 [D-F] \\
~~~~B20.sf2.uv2.bw40 & 0.75 & 0.2 & 40 & 5.7 & 3 [D-F] \\
~~~~B20LR.sf1.uv2.bw10 & 1.5  & 0.2 & 10 & 5.7 & 1 [B] \\
~~~~B20HR.sf1.uv2.bw10 & 1.5  & 0.2 & 10 & 5.7 & 1 [B] \\
\\
%\hline\\
\underline{$40h^{-1}\dim{Mpc}$ boxes} \\
\\
~~~~{\bf \bl}        & 1.5 & 0.1 & 10 & 5 & 3 [A-C] \\
~~~~B40.sf1.uv2.bw10 & 1.5 & 0.2 & 10 & 5.5 & 3 [A-C] \\
\\
\hline\\
\end{tabular}
\end{table}

Table \ref{tab:sets} lists all sets of simulations that we present in this paper. We have performed several sets of $20h^{-1}\dim{Mpc}$ that serve as our primary parameter exploration data. Most of the science results we extract from two sets of 3 independent realizations each of the $40h^{-1}\dim{Mpc}$ box. We highlight in boldface the simulation sets that we consider ``fiducial'' - in the next section we justify that particular choice. 

In order to distinguish individual realizations in each simulation set, we label them with capital letters. For example, the fiducial $20h^{-1}\dim{Mpc}$ set (\bm) includes six independent realizations A-F. A set B20.sf1.uv2.bw40 only includes 3 simulations D, E, and F, which have identical initial conditions to the simulations D, E, and F from the fiducial set \bm. Hence, we have the ability to both compare simulations with identical initial conditions but varied physical parameters (like \bm.D and B20.sf1.uv2.bw40.D) and simulations with identical physical parameters but different realizations of initial conditions (like \bm.A and \bm.B).

We also use comparison between $20h^{-1}\dim{Mpc}$ and $40h^{-1}\dim{Mpc}$ boxes as a rudimentary convergence test. Since every computed physical quantity may have different convergence requirements, we do not discuss numerical convergence in a separate sub-section, but rather include such discussion together with the calibration for each simulation parameter.

%---------------------------
\section{Calibration of the Simulations}
\label{sec:calib}
%---------------------------

\subsection{Star Formation Model}
\label{sec:lfun}

One of the largest existing observational data sets on the sources of reionization are the UV luminosity functions  of high redshift galaxies. Matching them would validate our star formation and feedback models.

As our final simulation data, we combine the fiducial $20h^{-1}\dim{Mpc}$ and $40h^{-1}\dim{Mpc}$ sets (\bm\ and \bl). Six independent realizations of the $20h^{-1}\dim{Mpc}$ box are equivalent, by volume, to 0.75 of a single $40h^{-1}\dim{Mpc}$ box, increasing our fiducial set from 3 to 3.75 $40h^{-1}\dim{Mpc}$ boxes.

\begin{figure}[t]
\includegraphics[width=\hsize]{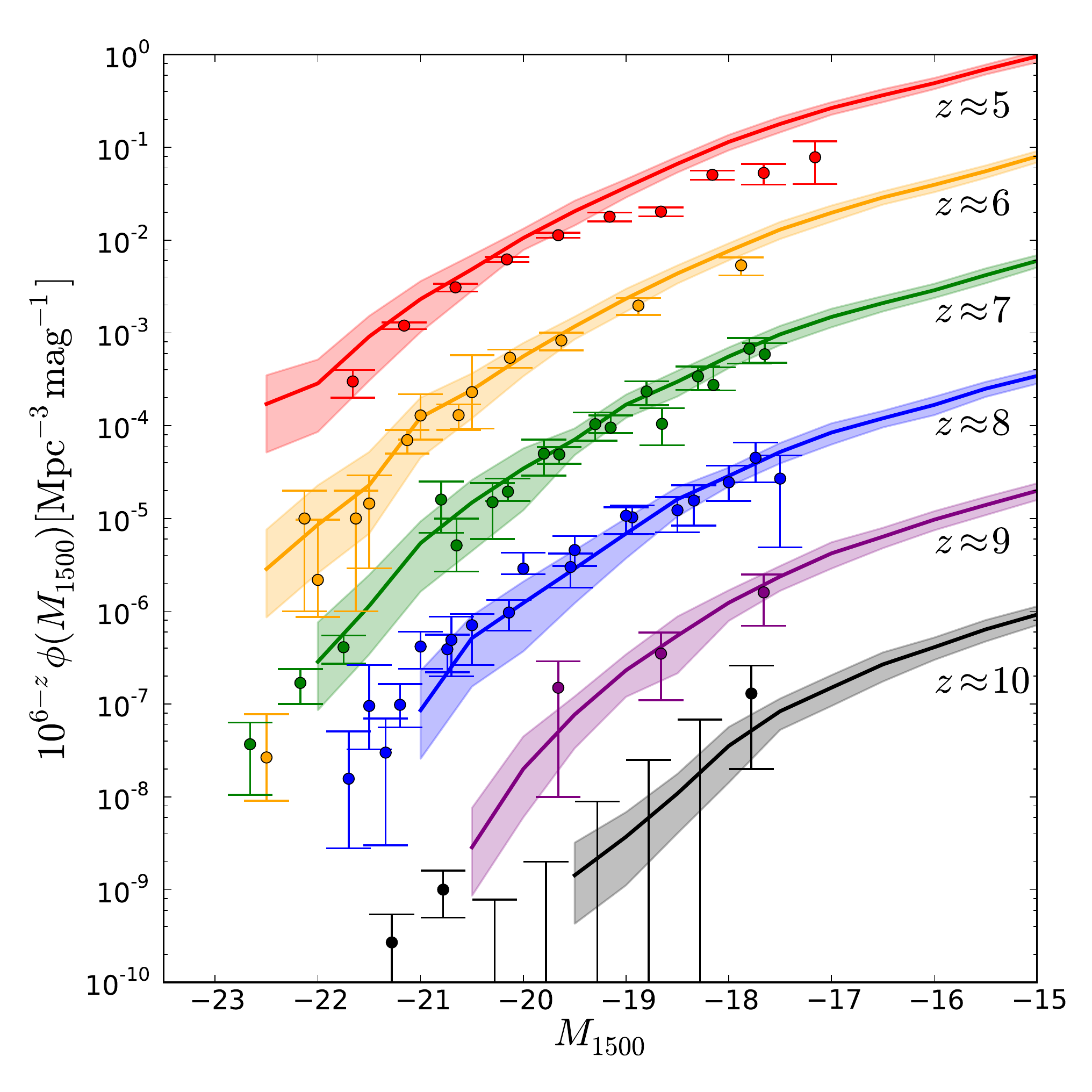}
\caption{Ultraviolet galaxy luminosity functions from a combination of simulation sets \bm\ and \bl\ at 6 different redshifts. Lines with bands show the average luminosity functions with the rms variation over our effective 3.75 $40h^{-1}\dim{Mpc}$ boxes (the error of the mean is, respectively, a factor of $\sqrt{3.75}=1.9$ smaller). Circles with error-bars are a compilation of recent observational measurements \citep{gals:biff07, gals:biol11,rei:obig12,rei:btos12,rei:sreo13,rei:wmhb13,rei:obil13,rei:bdm14,rei:obi14}. Different redshifts are shifted vertically by 1 dex for clarity.\label{fig:lfall}}
\end{figure}

In order to predict UV luminosities of our model galaxies, we use the Flexible Spectral Population Synthesis (FSPS) code \citep{misc:cgw10,misc:cg10}. One complication in computing stellar luminosities in far UV is a proper account of cosmic dust. A fully self-consistent dust model would require a complex and computationally expensive ray-tracing through the simulated galaxies, and is beyond the scope of this first paper. Instead, we adopt a simple but reasonable dust obscuration model, which we delegate to the appendix, as it has only modest effect on our results (and no effect at all at $z>7$).

Figure \ref{fig:lfall} presents the primary result of this paper - the evolution of the galaxy UV luminosity function between $z=10$ and $z=5$ from our simulations, and its comparison with the existing observational measurements\footnote{The exact values of redshifts are matched to Hubble filters used in observations: $z=5.0, 5.9, 6.8, 8.0, 9.0, 10.0$}. A simple star formation model of Equation (\ref{eq:sfr}) is able to reproduce the observations for all $z>5$. The agreement becomes worse at low luminosities at $z=5$, and that is not particularly surprising - since our simulations maintain the fixed resolution in comoving units, the spatial resolution degrades as simulations evolve, and the feedback model becomes progressively less accurate, especially in the low mass galaxies.

\begin{figure}[t]
\includegraphics[width=\hsize]{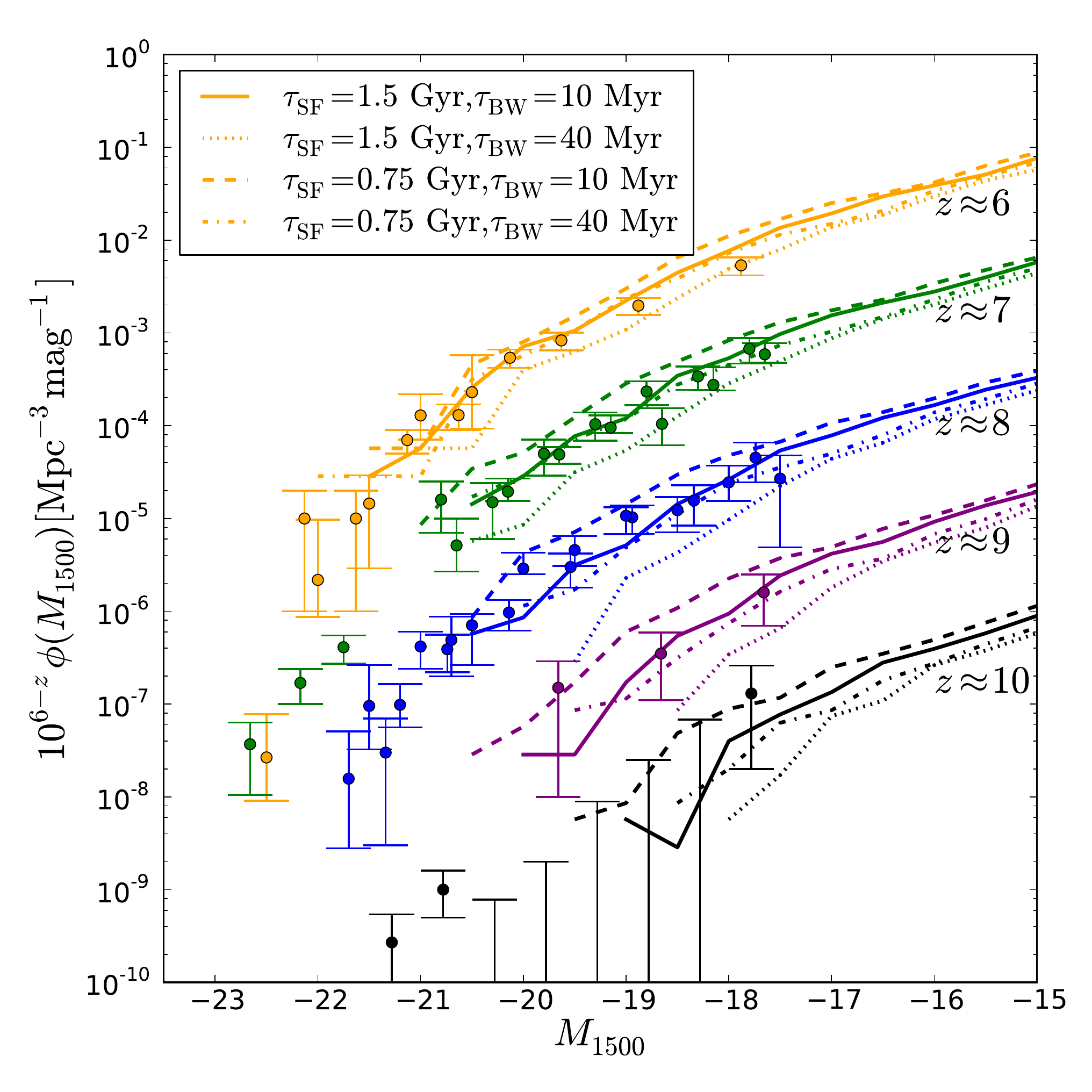}
\caption{Ultraviolet galaxy luminosity functions for simulation sets with varied parameters of the star formation model. The time-scale for delayed cooling $\tbw$ and the depletion time $\tsf$ (i.e.\ star formation efficiency) do affect the computed luminosity functions, but \emph{not} a directly proportional way due to well-known self-regulation of star formation.\label{fig:lfsfr}
}
\end{figure}

The sensitivity of our star formation model to the numerical parameters $\tsf$ and $\tbw$ is explored in Figure \ref{fig:lfsfr}. To compare apples and apples, we use only 3 independent realizations D-F from our fiducial set \bm, and compare them with simulation sets B20.sf1.uv2.bw40, B20.sf2.uv2.bw10, and B20.sf2.uv2.bw40 (which only included 3 simulations D-F each, and were only continued to $z=5.7$, to save computational resources). A longer time-scale $\tbw$ for delayed cooling does have a significant effect on the simulated galaxies, but \emph{not} a directly proportional one - luminosity functions for the set B20.sf1.uv2.bw40 match the fiducial set very well if shifted horizontally by about 0.5 magnitude, which corresponds to only a factor of 1.6. The sensitivity of our star formation model to the star formation time-scale $\tsf$ (or, equivalently, to star formation efficiency) is stronger -  a change in $\tsf$ by a factor of 2 makes a similar 0.5 magnitude change of the simulated luminosity functions, but the effect is still substantially sub-linear. Finally, since the two parameters have opposite effects, a higher star formation efficiency (lower $\tsf$) can always be compensated by a stronger feedback (longer $\tbw$), as is illustrated by the set B20.sf2.uv2.bw40. That result should not be surprising at all - it is well established that stellar feedback ``self-regulates'' star formation on kiloparsec scales \citep{sims:svbw09,sims:atm11,sims:hqm11,ng:akl13,sims:hko13}. The sensitivity of our star formation model to numerical parameters is somewhat higher than is usually found at lower redshifts, reflecting the fact that complete self-regulation takes some time to get established; that is also consistent with prior work \citep[e.g.][]{sims:svbw09,sims:atm11}.

\begin{figure}[t]
\includegraphics[width=\hsize]{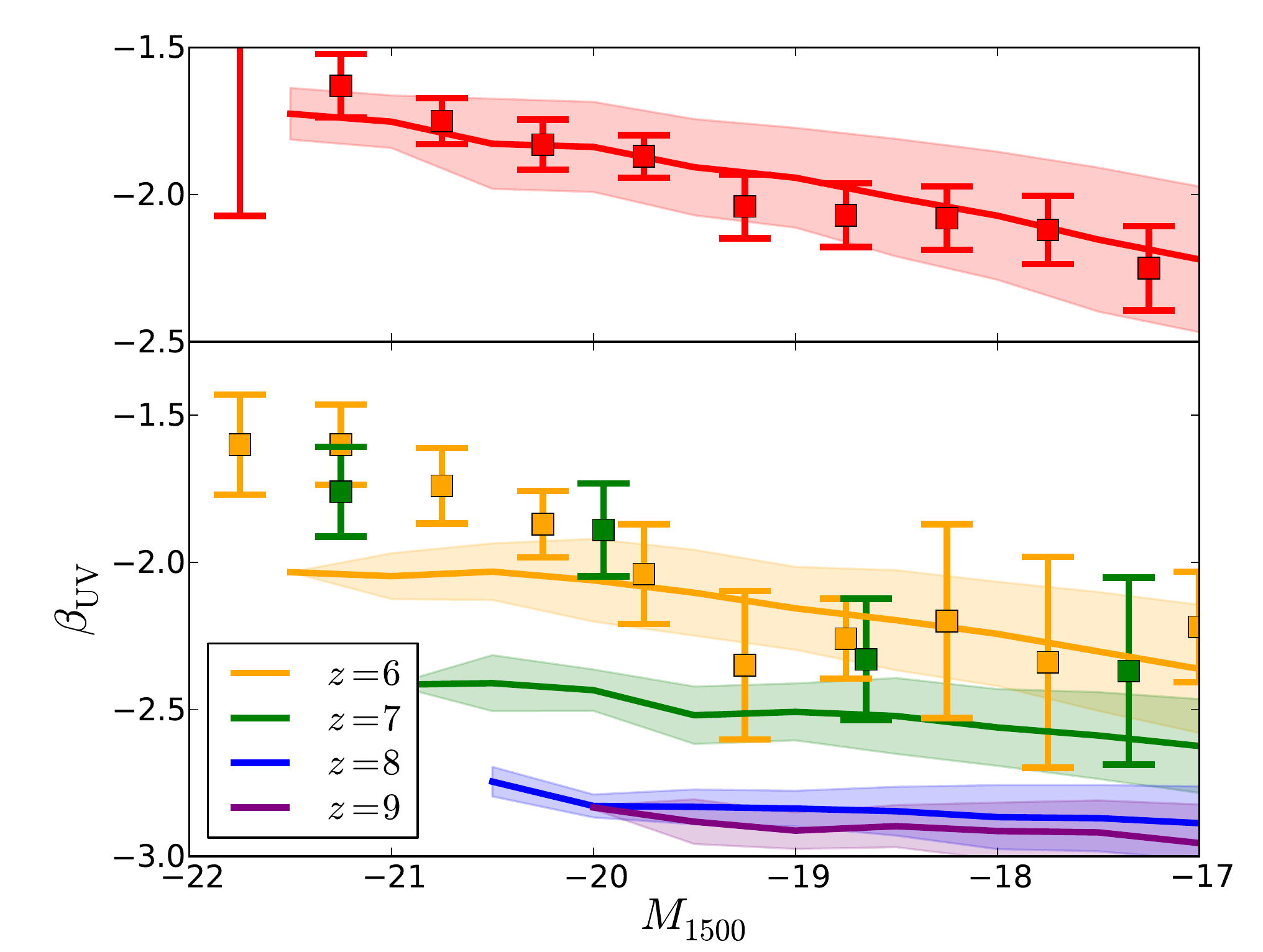}
\caption{
Ultraviolet continuum slopes of galactic SEDs for our simulated galaxies as a function of UV luminosity (average - lines, rms - shaded). Squares with error-bars are observational measurements from \protect\citet{rei:biol14}; $z=5$ is shown on a separate panel to reduce clutter in the figure. We do not match the slopes of the brightest galaxies at $z>5$, and the reason for the discrepancy is currently unclear.
\label{fig:beta}}
\end{figure}

Another observational constraint on our star formation model is offered by the observed UV continuum slopes of high redshift galaxies. To measure the UV continuum slopes of model galaxies, we compute the monochromatic fluxes at $1300\AA$, $1400\AA$, $1500\AA$, $1600\AA$, and $1700\AA$ for our fiducial \bm\ set\footnote{Such calculations would be too computationally expensive for all of \bl\ simulations, and are not really needed.}, and fit a power-law relation $F_\lambda \propto \lambda^\beta$ using a simple least-squares fit in a log-log plane. In Figure \ref{fig:beta} we show the average UV continuum slopes measured from our simulations and the rms scatter around them, as well as observational measurements at $z=5-7$. Our simulations match the observed slopes at $z=5$ rather well, but do not reproduce the trend of bluer spectra of brighter galaxies at $z=6$ and $z=7$. 

The latter discrepancy is somewhat surprising, since by matching the whole evolution of luminosity functions, our simulations reproduce not only star formation rates at a given redshift, but the whole prior star formation histories of galaxies. Never-the-less, it is possible to come up with several potential reasons for the discrepancy: our dust obscuration model (Appendix \ref{sec:dust}) may be over-simplistic (since it is independent of the galaxy luminosity), the FSPS code may not be sufficiently accurate, as may be observational, broad-band based diagnostics for the true spectral slope; finally, our star formation and stellar feedback models may be too crude. For all these reasons, we leave resolving this apparent discrepancy to future work.

\subsection{Ionizing Emissivity}
\label{sec:ionem}

Our star formation model, specified by parameters $\tsf$ and $\tbw$, appears to work reasonably well. The last remaining parameter in the simulations that needs to be calibrated is the escape fraction up to the simulation resolution limit $\euv$. This parameter can be calibrated with the observed spectra of high-redshift quasars - ionizing emissivity of our sources is proportional to $\euv$, and, therefore, the whole process of reionization and its aftermath - the \lya\ forest at $z=5-6$ - is affected by $\euv$.

In order to model absorption spectra of high-redshift quasars, we compute synthetic \lya\ spectra along 1000 totally random lines of sight at several simulation snapshots. Since simulations use periodic boundary conditions, we extend each line of sight to twice the simulation box size - random lines of sight in a periodic universe are only weakly affected by the artificial periodicity if they do not extend beyond twice the box length. For the $20h^{-1}\dim{Mpc}$ box that corresponds to about $\Delta z=0.15$ at $z=6$, which is a typical redshift resolution for averaging properties of \lya\ spectra in observed spectra \citep{igm:fsbw06}.

\begin{figure}[t]
\includegraphics[width=\hsize]{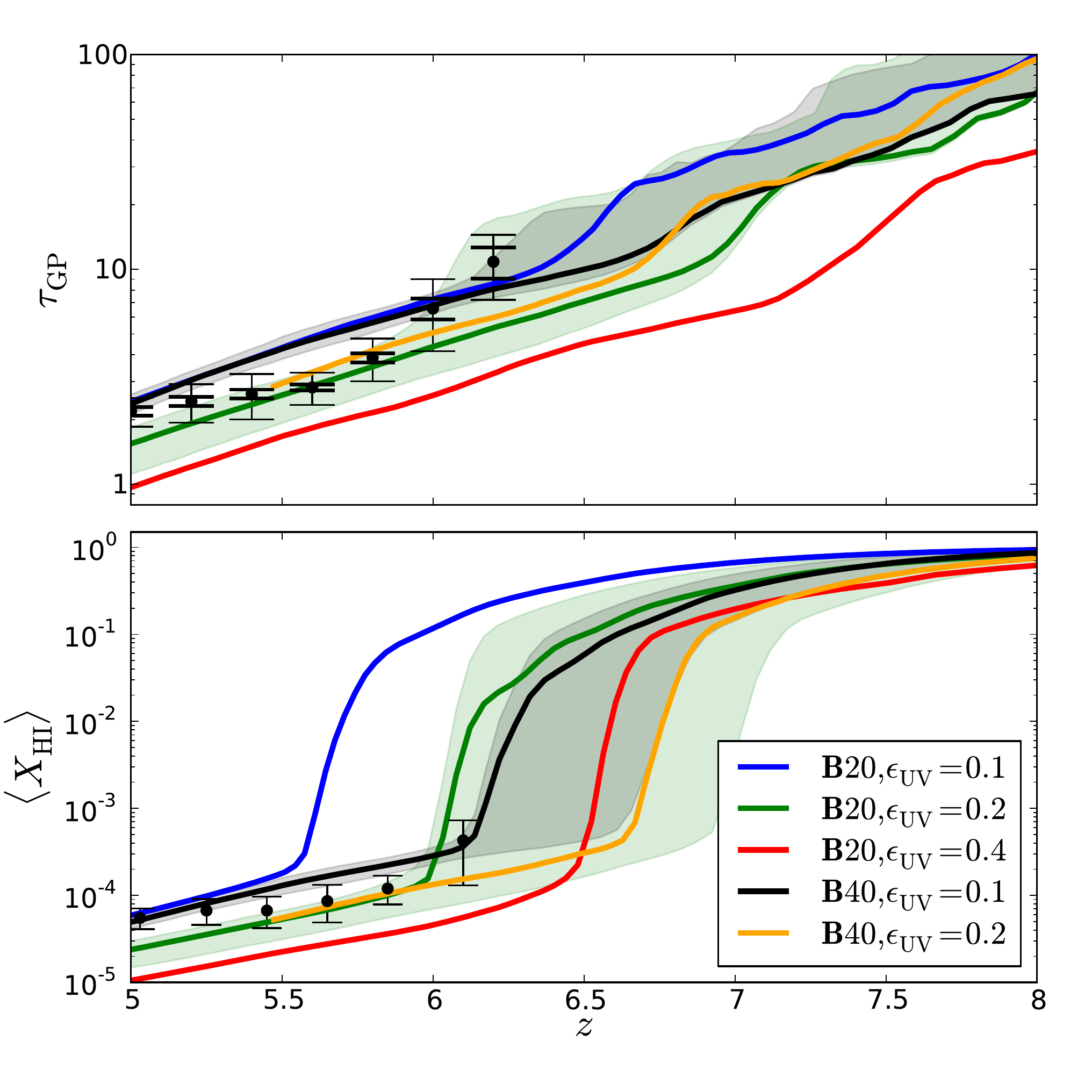}
\caption{Gunn-Peterson optical depth (top) and volume-weighted $\HI$ fraction (bottom) vs redshift for four simulation sets: B20.sf1.uv1.bw10 (blue), \bm\ (green), B20.sf1.uv4.bw10 (red), and \bl\ (black). Solid lines show the average quantities, while the shaded regions for fiducial sets \bm\ and \bl\ span the limits of variation among 6/3 independent realizations. Data points are from \protect\citet{igm:fsbw06}.\label{fig:xtz}}
\end{figure}

In Figure \ref{fig:xtz} we show the evolution of the mean Gunn-Peterson optical depth and the volume weighted mean $\HI$ fraction in four of our simulation sets: three $20h^{-1}\dim{Mpc}$ boxes with $\euv=0.1$, 0.2, and 0.4, and our fiducial $40h^{-1}\dim{Mpc}$ box. There are several conclusions that can be drawn from that figure.

Firstly, changing ionizing emissivity by a factor of 2 changes the moment of overlap of ionizing bubbles, indicated by the sharp drop in volume weighted mean $\HI$ fraction \citep{ng:g00a,ng:g04,ng:gf06}, by about $\Delta z\approx0.5$.

Secondly, the opacity of the post-reionization \lya\ forest are best matched by $\euv=0.2$ value independently of the box size.

Thirdly, and most importantly, we do not yet reach numerical convergence at the box size of $40h^{-1}\dim{Mpc}$ - the fact that out fiducial sets \bm\ and \bl\ have different values of the $\euv$ parameter, but similar reionization histories, indicates that even 3 independent realizations of a $40h^{-1}\dim{Mpc}$ are not enough to obtain an accurate prediction for the mean Gunn-Peterson optical depth or the volume weighted mean neutral fraction.

Finally, our simulations do not match the observational points at $z\ga6$ particularly well. However, we do not consider that a serious problem exactly for the third reason above: our sets of simulation boxes sample the \lya\ forest at each redshift interval way better than the observational measurements, and yet they are still far from convergence. Hence, the observational values are not the converged results either, and the discrepancy between the observational constraints and our simulations does not necessarily imply a major failure of our physical model, but may also be due to just cosmic variance.

\begin{figure}[t]
\includegraphics[width=\hsize]{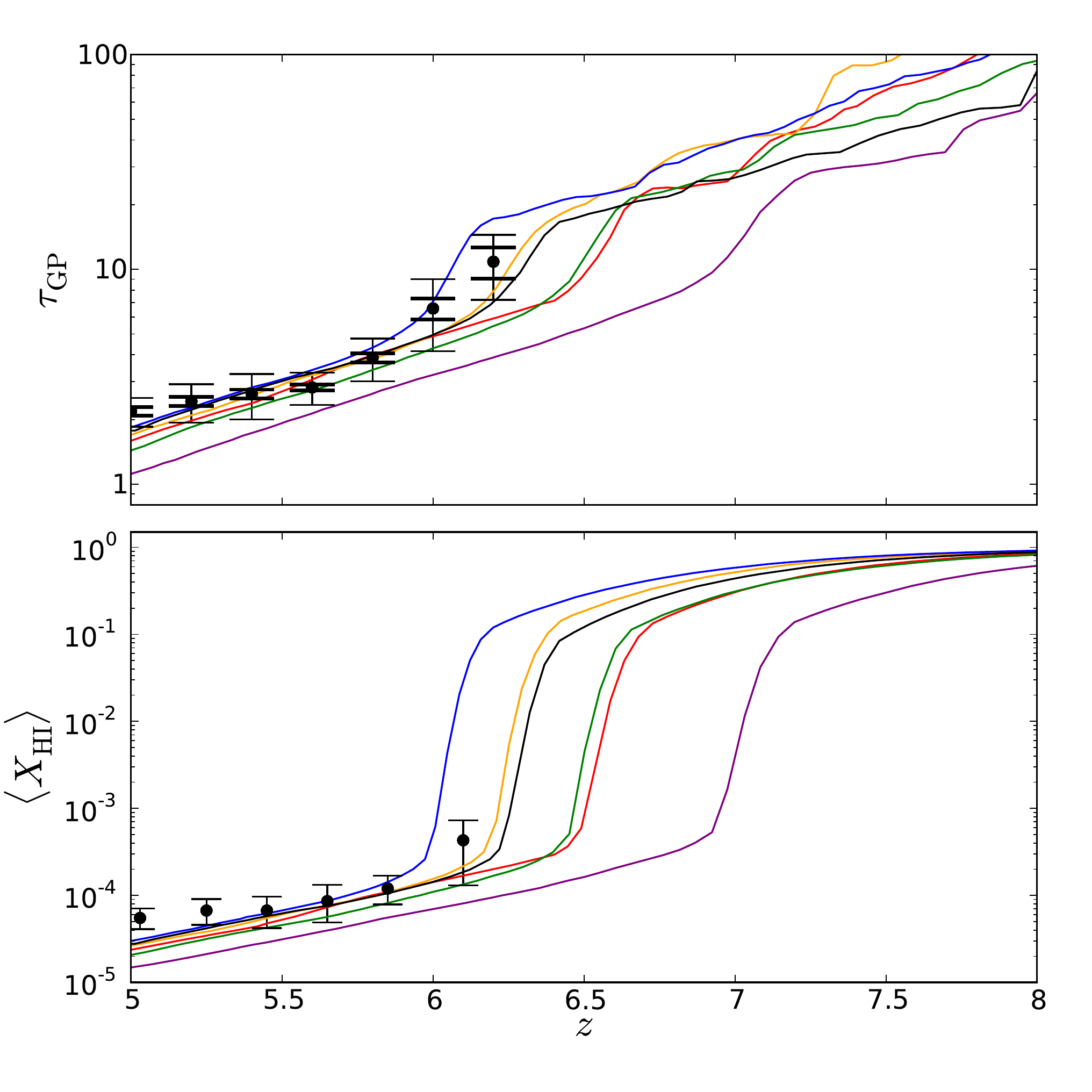}
\caption{Average Gunn-Peterson optical depth (top) and volume-weighted $\HI$ fraction (bottom) vs redshift for 6 independent realizations of the fiducial set \bm. Data points are from \protect\citet{igm:fsbw06}.\label{fig:xtz2}}
\end{figure}

Large cosmic variance is further illustrated in Figure \ref{fig:xtz2}, where we show the average Gunn-Peterson optical depth and the volume weighted mean $\HI$ fraction for all 6 independent realization of our fiducial \bm\ set. As one can see, the variations from realization to realization are very large (recall, that the $20h^{-1}\dim{Mpc}$ box is well-matched to the redshift bin of observational measurement of $\Delta z\approx 0.15$), and two of our realizations go through the observational points reasonably well.

The reason for such a large sensitivity to cosmic variance is in the extreme non-linearity of the relationship between the Gunn-Peterson optical depth $\tau_{\rm GP}$ and the gas density, which makes $\tau_{\rm GP}$ extremely sensitive to outliers. For example, consider $N$ lines of sight at some redshift during reionization, of which only one has non-zero transmitted flux $F_1$. The average flux over $N$ lines is $F_1/N$, and the corresponding Gunn-Peterson optical depth is 
\[
  \tau_{\rm GP} = -\ln\left(F_1/N\right) = \tau_1 + \ln N,
\]
where $\tau_1$ is the Gunn-Peterson optical depth along that one line of sight with partial absorption. In observations, typically only a few lines of sight contribute to a given redshift bin, i.e.\ $N=3-5$. Hence, a single line of sight with $\tau_1=10$ (an observational value at $z=6.2$) results in the measured ``average'' of $\tau_{\rm GP}\approx 11.1-11.6$, only slightly above $\tau_1$.

Based on this reasoning, we use only post-reionization \lya\ data ($z<6$) for calibrating our simulations, and rely on the derived volume-weighted mean $\HI$ fraction from \citet{igm:fsbw06} as preferred calibration data, even if the Gunn-Peterson optical depth is a directly observed quantity - $\langle X_\HI\rangle_V$ is, effectively, a convolution over the whole distribution function of observed \lya\ fluxes in individual spectral pixels, just like $\tau_{\rm GP}$, but it is more sensitive to typical values of the transmitted flux, while $\tau_{\rm GP}$ is heavily weighted towards the tail of the distribution. In that sense it is a less biased, even if less direct, observational probe.

Hence, our preferred value for the parameter $\euv$ is between 0.1 and 0.2 (with 0.2 matching observations better); the computational expense of simulations (and lack of convergence between $20h^{-1}\dim{Mpc}$ and $40h^{-1}\dim{Mpc}$ boxes) prevents us from actually fitting for the value of $\euv$ to higher precision.

%---------------------------
\section{Conclusions}
\label{sec:con}
%---------------------------

Cosmic Reionization On Computers (CROC) project is a long-term simulation campaign for modeling the process of cosmic reionization in sufficiently large simulations volumes (above 100 Mpc) with detailed physical modeling and spatial resolution better than 0.5 kpc (simulation cell size of less than 150 pc).

A simple model of star formation and feedback, based on the linear Kennicutt-Schmidt relation in the molecular gas and a widely used ``delayed cooling'' or ``blastwave'' feedback, is able to reproduce the observed galaxy UV luminosity functions in the whole redshift range $z=6-10$ with a value for the molecular gas depletion time of $\tsf=1.5\dim{Gyr}$, consistent with observations at $z\sim0$ and $z=1-2$.

A reasonable choice for the $\euv$ parameter, a quantity that describes photon losses on scales unresolved in our simulations, of $\euv=0.1-0.2$ results in reionization history that is reasonably consistent with the observed opacity of the \lya\ forest in the spectra of SDSS quasars at $z<6$. An even better consistency can be achieved by fitting the simulations to the data, albeit at (presently unrealistically) large computational expense.

The observed increase in the Gunn-Peterson optical depth at $z>6$ \citep{igm:fsbw06} has been interpreted by several groups (including ours) as evidence for the reionization overlap \citep{igm:bfws01,igm:wbfs03,ng:g04,igm:fck06,ng:gf06}. That increase is, however, a subject to large cosmic variance; with 6 independent realizations, each corresponding to multiple lines of sight, we find a spread in the redshift of overlap of about $\Delta z\approx1$. Since the observational constraints have even less statistical power than our simulations, they have not yet converged on the true evolution of the average Gunn-Peterson optical depth, and may, therefore, be significantly biased.

\acknowledgements

We are grateful to Andrea Ferrara and Matthew McQuinn for valuable comments and suggestions that significantly improved the original manuscript.

Simulations used in this work have been performed on the Joint Fermilab - KICP cluster ``Fulla'' at Fermilab, on the University of Chicago Research Computing Center cluster ``Midway'', and on National Energy Research Supercomputing Center (NERSC) supercomputers ``Hopper'' and ``Edison''.

\appendix

\section{Model for Quasar Sources}
\label{sec:qso}

For the intrinsic quasar SED we use our own fit to the \citet{qlf:rls06} model,
\[
  \nu L_\nu \propto \frac{9.5\times10^{-5}}{\left(1+\left(h\nu/300\dim{eV}\right)\right)^{0.8}}e^{\displaystyle -h\nu/500\dim{keV}} + 0.1\frac{\left(10\dim{eV}/h\nu\right)^{2.2}}{\left(1+\left(9\dim{eV}/h\nu\right)^5\right)^{0.4}} + 5.8e^{\displaystyle-h\nu/0.2\dim{eV}}.
\]

The evolution of the bolometric quasar luminosity function has been determined by \citet{qlf:hrh07}. We fit the quasar bolometric luminosity density as a function of redshift as
\[
  L_{\rm QSO}(z) = 10^{7.7}\frac{\Lsun}{\dim{Mpc}^3}\left(\left[e^{1.2(z-4)}\right]^{1/3} + \left[3e^{-2.7z}\right]^{1/3}\right)^{-3},
\]
and we use a $k$-correction of $k_{\rm ion}=3.9$ to translate from the bolometric to ionizing quasar luminosity.

\section{Dust Obscuration Model}
\label{sec:dust}

In this work we adopt a simple, but reasonable dust obscuration model for our simulated galaxies, in which the dust optical depth at wavelength $\lambda$ is estimated as
\begin{equation}
  \tau_{\lambda} = \sigma_D(\lambda) Z_* N_{\rm eff},
  \label{eq:dust}
\end{equation}
where $N_{\rm eff}$ is a parameter to be calibrated from the observational data, $Z_*$ is the stellar metallicity in solar units, and $\sigma_D(\lambda)$ is the dust cross-section at solar metallicity. For the specific extinction law, we take the SMC dust \citep{misc:wd01}, since most of our simulated galaxies at $z=5-6$ have metallicities similar to SMC, and normalize it to solar metallicity by assuming the SMC metallicity of -0.7 dex \citep[see][for details]{ng:gkc08},
\[ 
  \sigma_D(\lambda)=1.76\times10^{-21}\dim{cm}^2\left(\frac{1500\AA}{\lambda}\right)^{1.1}.
\]

\begin{figure}[t]
\includegraphics[width=0.5\hsize]{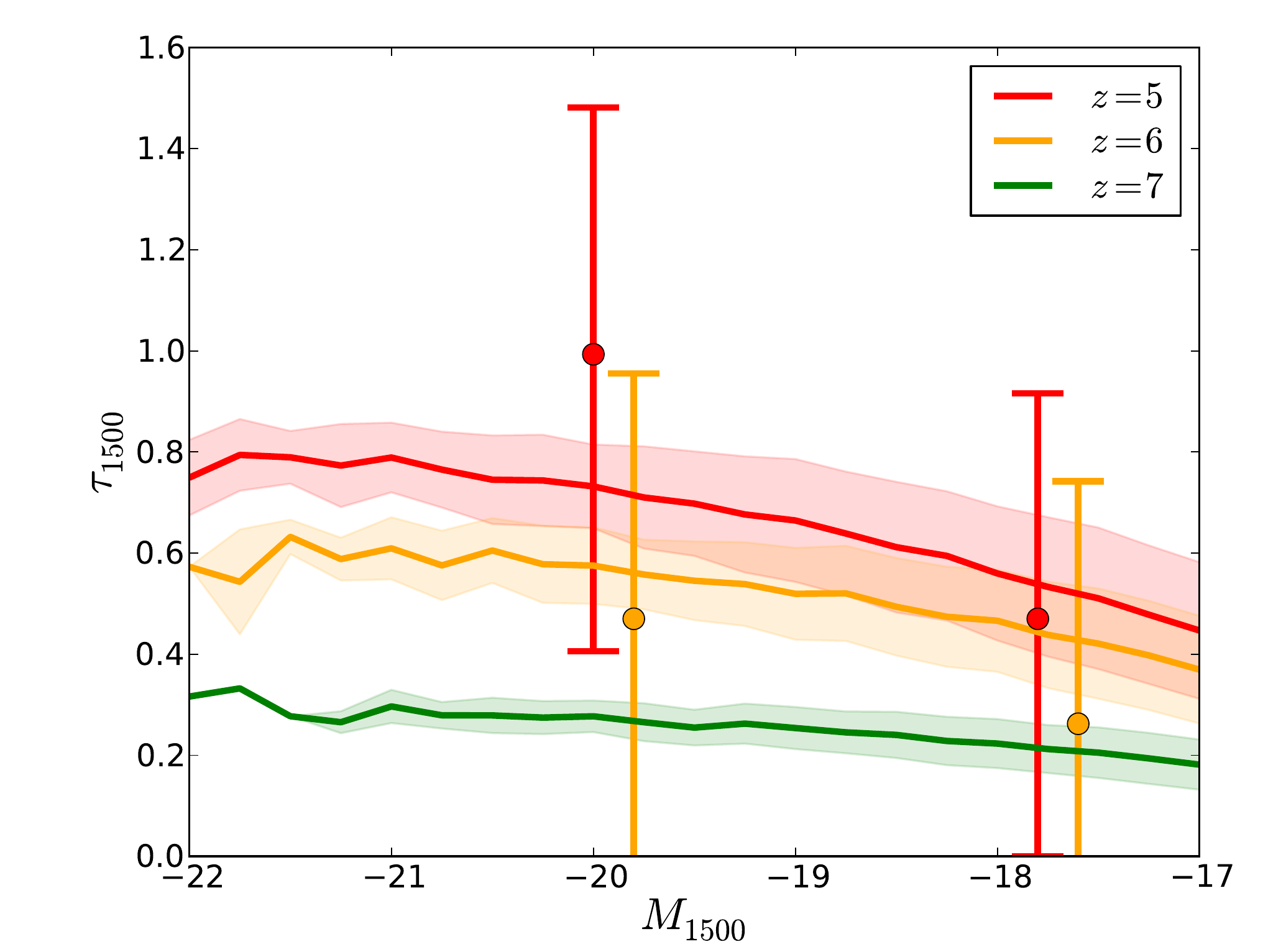}
\caption{Dust optical depth at $\lambda=1500\AA$ vs galaxy AB magnitude at the same wavelength. Points with error-bars are the observational constraints from \protect\citet{gals:bifc09}, while lines with bands are a combination of simulation sets \bm\ and \bl. Different colors correspond to redshifts in the rainbow order.\label{fig:dust}}
\end{figure}

Figure \ref{fig:dust} shows the observational constraints on the dust abundance in $z=5$ and $z=6$ galaxies from \citet{gals:bifc09}. Our simple dust obscuration model can roughly match these constraints with the value of $N_{\rm eff}\approx 5\times10^{21}\dim{cm}^{-2}$. At $z>7$ the observational constraints are consistent with no dust obscuration \citep{gals:bifc09,gals:biol11}; to account for that we adopt an extremely crude but simple ansatz for $N_{\rm eff}$:
\[
  N_{\rm eff} = 5\times10^{21}\dim{cm}^{-2}\min\left(1,\max\left(0,0.5(8-z)\right)\right)
\]
($N_{\rm eff}=5\times10^{21}\dim{cm}^{-2}$ for $z\leq6$, $N_{\rm eff}=0$ for $z\geq8$, and is linearly interpolated in between to avoid discontinuities). With that ansatz the dust obscuration at $z=7$ is not identically zero, but is small enough to be unimportant and undetectable observationally.

\section{Optically Thin Variable Eddington Tensor Approximation in the ART Code}
\label{sec:otvet}

\subsection{Two-field Ansatz for the Radiation Field}

Consider a radiative transfer equation in the expanding universe, in comoving reference frame,
\begin{equation}
  \frac{a}{c}\frac{\partial J_\nu}{\partial t} + \vec{n}\frac{\partial J_\nu}{\partial \vec{x}} - \frac{aH}{c} \left(\nu\frac{\partial J_\nu}{\partial \nu} - 3 J_\nu\right) = -k_\nu J_\nu + S_\nu,
  \label{eq:rt}
\end{equation}
where $J_\nu(t,\vec{x},\vec{n})$ is the radiation specific intensity, $k_\nu$ is the absorption coefficient (per unit comoving distance), and $S_\nu$ is the source function (in appropriate units).

In the following, it is assumed that
\begin{description}
\item[A1] all sources have the same spectral shape, i.e.\ the frequency dependence of $S_\nu$ can be factored out, and that
\item[A2] sources are isotropic ($S_\nu$ does not depend on $\vec{n}$).
\end{description}
Both these assumptions can be relaxed, if necessary. 

With these assumptions, we can write
\[
  S_\nu(t,\vec{x}) = L_\nu \rho_S(t,\vec{x}).
\]
For example, for stellar sources, $\rho_S(t,\vec{x})$ can be the mass density of massive stars.

In cosmological simulations, it is convenient to replace specific
intensity $J_\nu$ with two separate functions, $f_\nu$ and $g_\nu$, as
follows:
\begin{equation}
  J_\nu = \bar{J}_\nu f_\nu  + L_\nu (g_\nu - \bar{g}_\nu f_\nu),
  \label{eq:ansatz}
\end{equation}
where $\bar{J}_\nu$ is the spatially and angle averaged specific intensity (i.e.\ cosmic background), which satisfies the following equation,
\[
  \frac{a}{c}\frac{\partial \bar{J}_\nu}{\partial t} - \frac{aH}{c} \left(\nu\frac{\partial \bar{J}_\nu}{\partial \nu} - 3 \bar{J}_\nu\right) = - \bar{k}_\nu \bar{J}_\nu + \bar{S}_\nu,
\]
where we defined the mean absorption coefficient as radiation-field-weighted,
\[
  \bar{k}_\nu \equiv \frac{\langle k_\nu J_\nu\rangle}{\bar{J}_\nu}.
\]
Analogously, $\bar{g}_\nu$ is the spatially and angle averaged $g_\nu$.
 
The reason to impose such an ansatz is to simplify the frequency dependence of the radiation field - in principle, one would need to follow several hundred radiation fields, one for each frequency bins, in order to compute accurately most of ionization and other chemical rates. This is not practical, obviously. Hence, the goal of ansatz (\ref{eq:ansatz}) is to concentrate most of the frequency dependence in the pre-factors, and hope that both $f_\nu$ and $g_\nu$ depend on the frequency only moderately. 

For example, one can imagine that the radiation field shining on a particular place in space is a combination of a contribution of the cosmic background (perhaps, attenuated by additional local absorption) and the radiation from nearby sources (perhaps, also attenuated by additional local absorption). In that case $f_\nu$ and $g_\nu$ would only have to account for the local absorption, and in places where the local absorption is negligible, would become completely frequency independent.

The complete details of our approach to modeling the frequency dependence are given in \citet{ng:ga01}. Here we just mention that we represent the full frequency dependence of $f_\nu$ and $g_\nu$ as
\[
  f_\nu = f_0 \exp\left(-\sum_j q_j(\nu) w_j \right),
\]
where $q_j(\nu)$ are a set of pre-determined basis functions (not necessarily mutually orthogonal), and $w_j$ are weights, which need to be computed by the RT solver (the expression for $g_\nu$ is completely analogous). In the present implementation we choose $q_j(\nu)$ to be photo-ionization cross-sections for $\HI$, $\HeI$, and $\HeII$ (i.e. the sum over $j$ includes just 3 terms), and weights $w_j$ and the normalization $f_0$ (or $g_0$) are derived by sampling $f_\nu$ (or $g_\nu$) at $\HI$, $\HeI$, and $\HeII$ ionization thresholds as well in the optically thin limit. The advantage of this approach is that in the most likely physical situation - when the opacity at a given location is dominated by a single system along a line of sight to a radiation source (or to the radiation background in case of $g_\nu$) - the full frequency dependence of the incident radiation field is recovered exactly.

The specific form of ansatz (\ref{eq:ansatz}) is dictated by the need to satisfy the consistency condition, since $J_\nu$ now depends on its own average; averaging both sides of Equation (\ref{eq:ansatz}) over space and angles gives
\[
  \bar{J}_\nu = \bar{J}_\nu \bar{f}_\nu + L_\nu (\bar{g}_\nu - \bar{g}_\nu \bar{f}_\nu),
\]
which can be satisfied if $\bar{f}_\nu = 1$.

There are more than one way to introduce the ansatz similar to Equation (\ref{eq:ansatz}). That particular form, however, ensures that both $f_\nu$ and $g_\nu$ remain always non-negative, which is a desirable property for a numerical implementation (as it avoids numerical loss of precision).

Equation (\ref{eq:ansatz}) is nothing more than an ansatz, we replaced one unknown function with two, hence we can impose a condition on these two functions. The condition we impose is that terms with $f_\nu$ and $g_\nu$ cancel out separately. In the following we drop the frequency subscripts for brevity, all quantities except $\rho_S$ remain functions of frequency. In addition, we adopt a Newtonian approximation for $f$ and $g$, omitting terms with the Hubble parameter for them, but retaining cosmological terms for $\bar{J}_\nu$ \citep[see][for more detailed description of this approximation]{ng:ga01}. With these simplifications, one obtains
\[
  f\left[-\bar{k}\bar{J}+\bar{S}\right] - Lf\frac{a}{c}\frac{\partial \bar{g}}{\partial t} + (\bar{J}-L\bar{g}) \frac{Df}{dl} + L \frac{Dg}{dl} =
\]
\begin{equation}
  = -(\bar{J}-L\bar{g}) k f - L k g + L \rho_S,
  \label{eq:fgmaster}
\end{equation}
where we use a shorthand
\[
  \frac{D}{dl} \equiv \frac{a}{c}\frac{\partial}{\partial t} + \vec{n}\frac{\partial}{\partial \vec{x}}
\]
for the derivative along the light cone. 

The condition we impose is then
\begin{equation}
  \frac{Dg}{dl} = -kg + \rho_S,
  \label{eq:g}
\end{equation}
which, in Newtonian limit (ignoring terms with $1/c$) has a simple solution,
\begin{equation}
  g(t,\vec{x},\vec{n}) = \int_0^\infty dl\, \rho_S(t,\vec{x}+\vec{n}l) e^{-\tau(\vec{x},\vec{x}+\vec{n}l)},
  \label{eq:gsol}
\end{equation}
where $\tau(\vec{x}_1,\vec{x}_2)$ is the optical depth between points $\vec{x}_1$ and $\vec{x}_2$,
\[
  \tau(\vec{x}_1,\vec{x}_2) = |\vec{x}_1-\vec{x}_2|\int_0^1 ds\, k\left(\vec{x}_1+s(\vec{x}_1-\vec{x}_2)\right).
\]
An even more familiar form is the angle average of $g$,
\[
  \langle g \rangle(t,\vec{x}) = \frac{1}{4\pi} \int d^3x^\prime \frac{\rho_S(\vec{x^\prime})}{(\vec{x}-\vec{x^\prime})^2} e^{-\tau(\vec{x},\vec{x^\prime})},
\]
which is simply an integral of $\rho_S/(4\pi r^2)$ over all sources, diminished by the opacity between the source and the current location. In particular, $g$ is manifestly positive everywhere in the computational domain.

Using Equation (\ref{eq:g}) in (\ref{eq:fgmaster}), we find
\[
  (\bar{J}-L\bar{g}) \left[ \frac{Df}{dl} + kf \right] =   
  f\left[\bar{k}\bar{J}-L\bar{\rho_S}\right] + Lf\frac{a}{c}\frac{\partial \bar{g}}{\partial t}.
\]
Averaging Equation (\ref{eq:g}) over space and angle results in
\[
  \frac{a}{c}\frac{\partial \bar{g}}{\partial t} = -\langle kg \rangle + \bar{\rho_S}.
\]
Combining the last two equations together, we find
\begin{equation}
  (\bar{J}-L\bar{g}) \left[ \frac{Df}{dl} + kf \right] =   
  f\left[\bar{k}\bar{J}-L\langle kg \rangle\right].
  \label{eq:fint}
\end{equation}
Finally, we can use Equation (\ref{eq:ansatz}) to compute the average absorption,
\[
  \bar{k}\bar{J} \equiv \langle kJ \rangle = (\bar{J}-L\bar{g}) \langle kf \rangle + L \langle kg \rangle.
\]
Substituting $\langle kg \rangle$ from the above equation into Equation (\ref{eq:fint}), we obtain the final equation for the function $f$,
\begin{equation}
  \frac{Df}{dl} = -kf + f\langle kf \rangle.
  \label{eq:fu}
\end{equation}
Equations (\ref{eq:g}) and (\ref{eq:fu}) are our master equations for the cosmological radiative transfer. At this point no approximations have been made except for the assumptions A1 and A2 above.

One undesirable property of Equation (\ref{eq:fu}) is that it numerically unstable. To see that, we can average it over space and angle,
\begin{equation}
  \frac{a}{c}\frac{\partial \bar{f}}{\partial t} = \langle kf \rangle (\bar{f} - 1).
  \label{eq:fbar}
\end{equation}
Value $\bar{f}=1$ is indeed a solution of this equation, but an unstable one: $\frac{\partial \bar{f}}{\partial t}>0$ for $\bar{f}>1$ and $\frac{\partial \bar{f}}{\partial t}<0$ for $\bar{f}<1$. To circumvent this problem, we multiply the last term in equation (\ref{eq:fu}) by a function $q(\bar{f})$,
\begin{equation}
  \frac{Df}{dl} = -kf + q(\bar{f})f\langle kf \rangle,
  \label{eq:f}
\end{equation}
where $q(1) = 1$. It is easy to show that, if $q^\prime(1) < -1$, then Equation (\ref{eq:f}) is numerically stable.

A fiducial choice for $q$ is
\[
  q(x) = \frac{2}{x} - 1,
\]
but, in the future, other forms for that function need to be explored.

\subsection{OTVET Approximation in the Two-field Ansatz}

In \citet{ng:ga01} it is shown how to derive a single diffusion-like equation for the angle average of fields $f$ and $g$. Namely, if
\begin{eqnarray}
  F_\nu(t,\vec{x}) & \equiv & \frac{1}{4\pi} \int d\Omega f_\nu(t,\vec{x},\vec{n}), \nonumber \\
  G_\nu(t,\vec{x}) & \equiv & \frac{1}{4\pi} \int d\Omega g_\nu(t,\vec{x},\vec{n}), \nonumber 
\end{eqnarray}
then (again omitting the frequency dependence for brevity)
\begin{eqnarray}
  \frac{\partial G}{\partial \xi} & = & \frac{\partial}{\partial x^j}\left(\frac{1}{k}\frac{\partial G h_G^{ij}}{\partial x^i}\right) - kG + \rho_S, \label{eq:G} \\
  \frac{\partial F}{\partial \xi} & = & \frac{\partial}{\partial x^j}\left(\frac{1}{k}\frac{\partial F h_F^{ij}}{\partial x^i}\right) - kF + q(\bar{F})F\langle kF \rangle, \label{eq:F} 
  \label{eq:FG}
\end{eqnarray}
where $d\xi = \hat{c}\,dt/(2 a)$, $\hat{c} \leq c$ is the ``reduced speed of light'' \citep{ng:ga01}, and averaging in Equation (\ref{eq:F}) is done over the space (obviously, $\bar{f} = \bar{F}$).

We choose Eddington tensors differently in Equations (\ref{eq:G}) and (\ref{eq:F}): for a cosmological simulation in a periodic box, $h_G^{ij}$ is chosen as the optically thin Eddington tensor from all sources inside a periodic box, while the Eddington tensor for the background radiation is taken to be isotropic, $h_F^{ij} = \delta^{ij}/3$.

\subsection{Elliptic Solver for OTVET Diffusion-like Equation}

Consider OTVET-type equation for some function $E(\xi,\vec{x})$ and some tensor $h^{ij}(\xi,\vec{x})$,
\begin{equation}
  \frac{\partial E}{\partial \xi} = \frac{\partial}{\partial x^j}\left(\frac{1}{k}\frac{\partial Eh^{ij}}{\partial x^i}\right) - kE + s,
  \label{eq:E}
\end{equation}
where $k(\xi,\vec{x})$ is the absorption coefficient and $s(\xi,\vec{x})$ is the source term. This equation is discretized in some fashion in space on a set of indicies $\{I\}$, where $I$ may may an SPH particle number, a set of indicies $(i,j,k)$ on a regular mesh, or any other discretization. We assume that the discretization is such that all quantities $E_I$, $h_I^{ij}$, $k_I$, and $s_I$ are co-located in space on the same set of resolution elements $\{I\}$.

The discretization scheme is associated with a spatial scale $\Delta x$ (cell size on a regular mesh, SPH kernel size, etc). Each cycle of the OTVET solver (for example, a time-step of a hydro scheme) consists of the set of consecutive iterations, which we label as $E_I^{(n)}$, where $n=0, 1, 2, ...$. 

The second-order term in Equation (\ref{eq:E}) is discretized as
\[
 \left.\frac{\partial}{\partial x^j}\left(\frac{1}{k}\frac{\partial E h^{ij}}{\partial x^i}\right)\right|_I \approx
\frac{1}{\Delta x} \hat{\cal D}_I[E;a],
\]
where $a_I \equiv k_I \Delta x$ is the dimensionless absorption coefficient and $\hat{\cal D}_I[E;a]$ is a linear operator on the set of all values $E_I$,
\[
  \hat{\cal D}_I[E;a] = \sum_J w_{I,J} E_J,
\]
with the sum being over all indicies $J$. Dimensionless weights $w_{I,J}$ depend on various $a_I$ as appropriate, and most of $w_{I,J}$ are zero because the second order term only has a finite support.

For example, for a regular mesh with $I=(i,j,k)$,
\[
  \hat{\cal D}_I[E;a] = \frac{U^x_{i+1,j,k}-U^x_{i,j,k}}{a_{i+1/2,j,k}} + \frac{U^x_{i-1,j,k}-U^x_{i,j,k}}{a_{i-1/2,j,k}} + \frac{U^y_{i,j+1,k}-U^y_{i,j,k}}{a_{i,j+1/2,k}} + \frac{U^y_{i,j-1,k}-U^y_{i,j,k}}{a_{i,j-1/2,k}} + \frac{U^z_{i,j,k+1}-U^z_{i,j,k}}{a_{i,j,k+1/2}} + \frac{U^z_{i,j,k-1}-U^z_{i,j,k}}{a_{i,j,k-1/2}},
\]
where $a_{i+1/2,j,k} = (a_{i,j,k}+a_{i+1,j,k})/2 + \epsilon$, etc, and a small offset $\epsilon=10^{-3}$ is added to avoid division by zero (we have verified it makes no effect on the actual solution). Fluxes $U^j$ are defined as
\begin{eqnarray}
  U^x_{i,j,k} & = & E_{i,j,k}h^{xx}_{i,j,k} + \frac{1}{4}\left(E_{i,j+1,k}h^{xy}_{i,j+1,k}+E_{i,j-1,k}h^{xy}_{i,j-1,k}+E_{i,j,k+1}h^{xz}_{i,j,k+1}+E_{i,j,k-1}h^{xz}_{i,j,k-1}\right), \nonumber \\
  U^y_{i,j,k} & = & E_{i,j,k}h^{yy}_{i,j,k} + \frac{1}{4}\left(E_{i+1,j,k}h^{yx}_{i+1,j,k}+E_{i-1,j,k}h^{yx}_{i-1,j,k}+E_{i,j,k+1}h^{yz}_{i,j,k+1}+E_{i,j,k-1}h^{yz}_{i,j,k-1}\right), \nonumber \\
  U^z_{i,j,k} & = & E_{i,j,k}h^{zz}_{i,j,k} + \frac{1}{4}\left(E_{i+1,j,k}h^{zx}_{i+1,j,k}+E_{i-1,j,k}h^{zx}_{i-1,j,k}+E_{i,j+1,k}h^{zy}_{i,j+1,k}+E_{i,j-1,k}h^{zy}_{i,j-1,k}\right), \nonumber 
\end{eqnarray}
etc.

Let us define two new, iteration-independent, discretized quantities,
\[
  A_I \equiv \frac{\gamma}{1+\gamma\left(\beta a_I - w_{I,I}\right)}
\]
($w_{I,I}<0$ is the diagonal term of the operator $\hat{\cal D}_I[E;a]$) and
\[
  B_I \equiv s_I \Delta x - (1-\beta) a_I E_I^{(0)},
\]
where $\alpha$, $\beta$, and $\gamma$ are constants.

Then one iteration of the OTVET elliptic solver consists of computing
\[
  d^{(n)}_I = \hat{\cal D}_I[E^n;a] - \beta a_I E_i^n + B_I,
\]
for all $I$, followed by updating $E_I$ as
\[
  E^{(n+1)}_I = E^{(n)}_I + \alpha A_I d^{(n)}_I.
\]
Numerical stability requires $\alpha<1$. A set of values that works particularly well is 
\[
  \begin{array}{ccc}
  \alpha & = & 0.8,\\
  \beta  & = & 0.1,\\
  \gamma & = & 1.\\
  \end{array}
\]

\begin{figure}[t]
  \includegraphics[width=0.5\hsize]{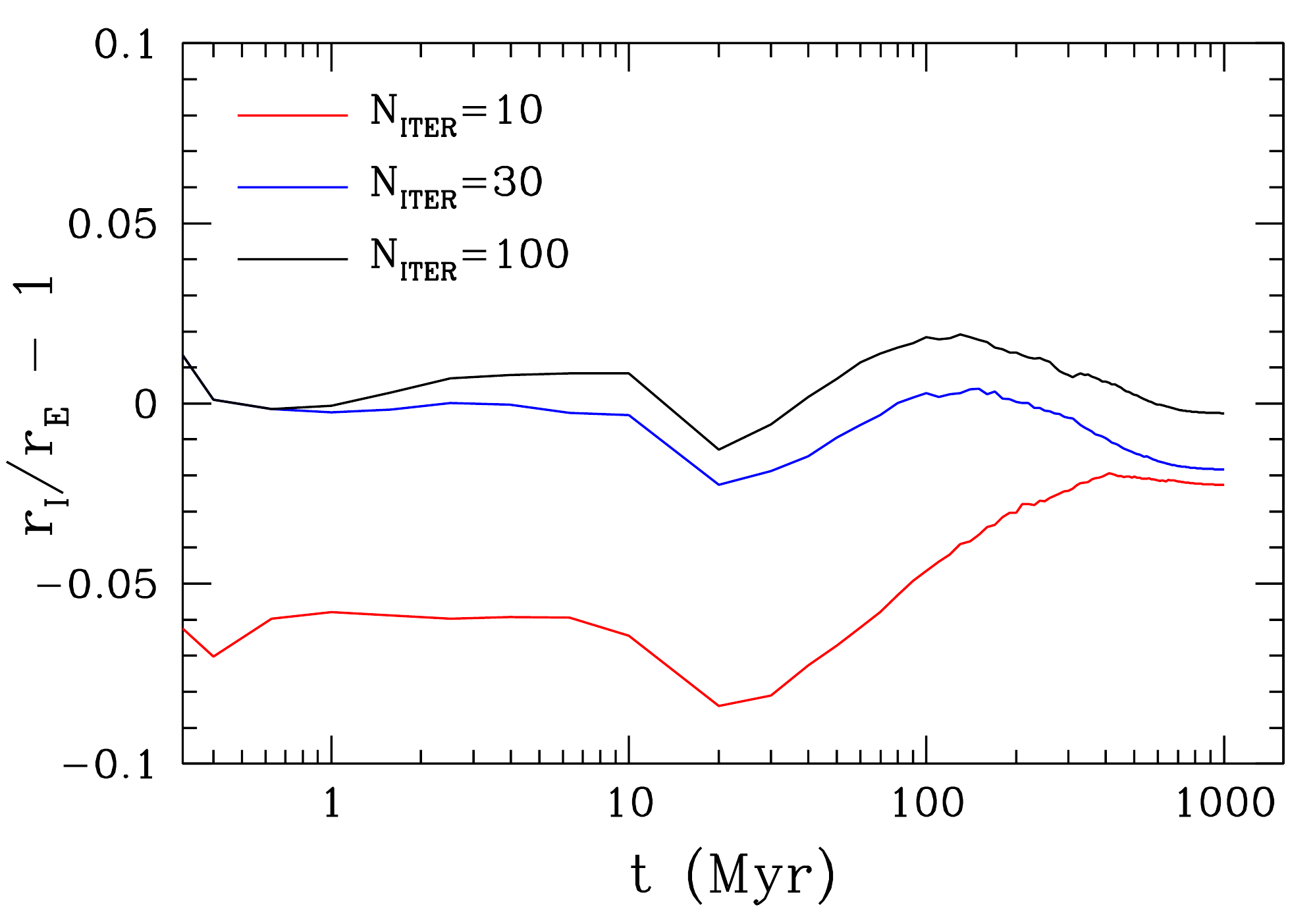}
  \includegraphics[width=0.5\hsize]{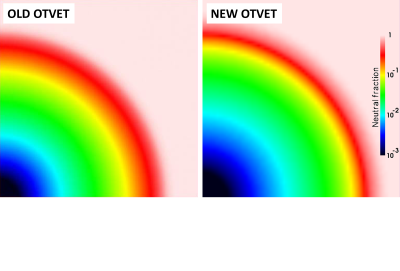}
  \caption{\label{fig:iliev1} Left: Error in the position of the ionization front $r_I$ as a function of time in the   \citet{ng:icam06} Test \# 1 for several numbers of iterations (as indicated on the legend). About 30 iterations are sufficient to achieve the convergence. Right: Comparison of the old and new OTVET implementations in Test 1 of \protect\citet{ng:icam06}. In the new implementation the I-front is sharper and it location is more accurate. This images are to be compared with Figure 6 of \protect\citet{ng:icam06}.}
\end{figure}

The left panel of Figure \ref{fig:iliev1} shows the error in the propagation of the ionization front as a function of time in the Test \#1 of \citet{ng:icam06}. This is the most sensitive to the iteration count of all tests presented in \citet{ng:icam06} and \citet{ng:icam09}. As one can see, about 30 iterations are sufficient to achieve 2\% precision in the ionization front evolution, and mere 10 iterations already give 5\% precision. The right panel of the same figure shows the improvement in tracking the ionization front with the new scheme. It should be compared with Figure 6 of \citet{ng:icam06} - now the quality of OTVET solution approaches that of the best ray-tracing schemes.

\section{Mass Convergence Test}
\label{sec:massconv}

\begin{figure}[t]
  \includegraphics[width=0.5\hsize]{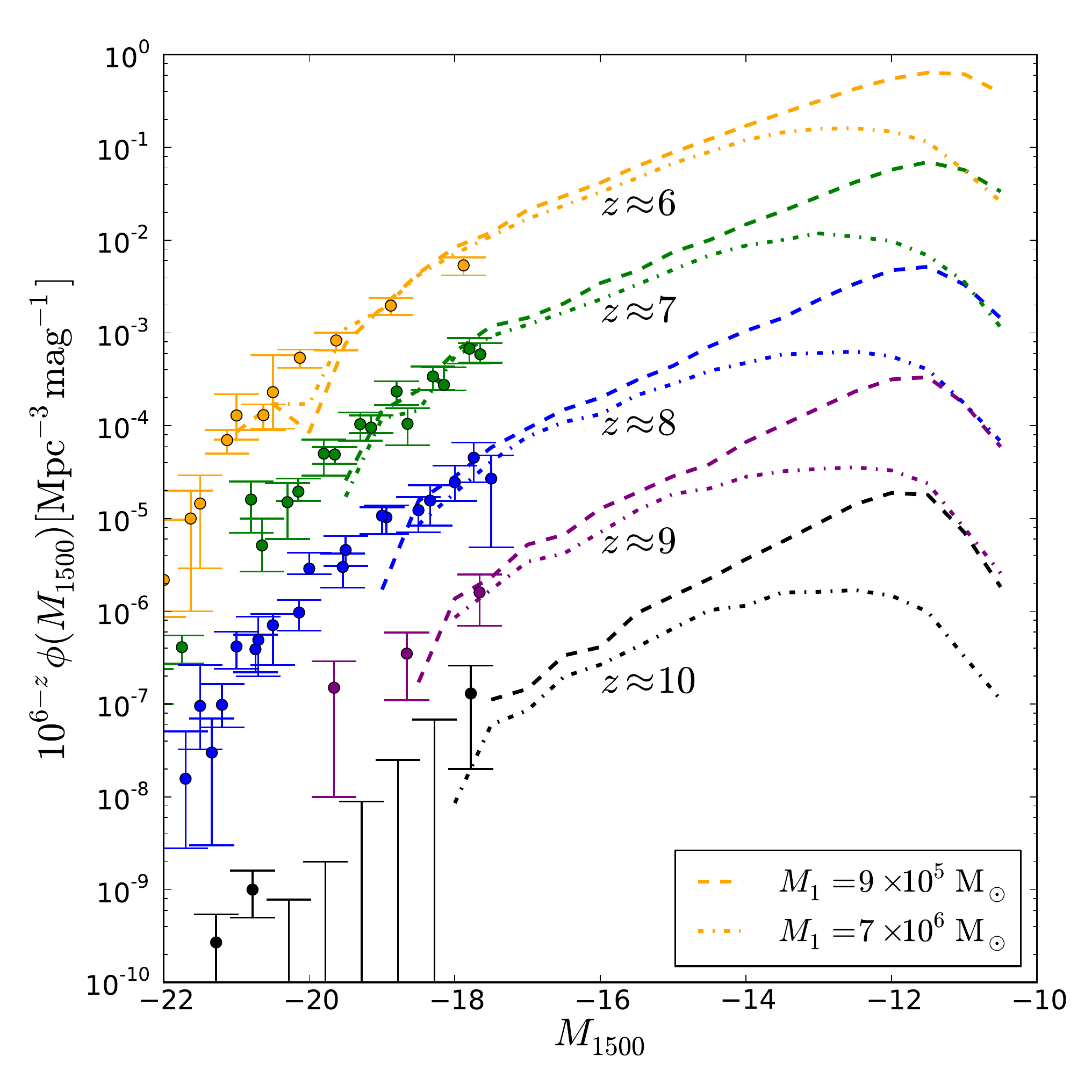}
  \includegraphics[width=0.5\hsize]{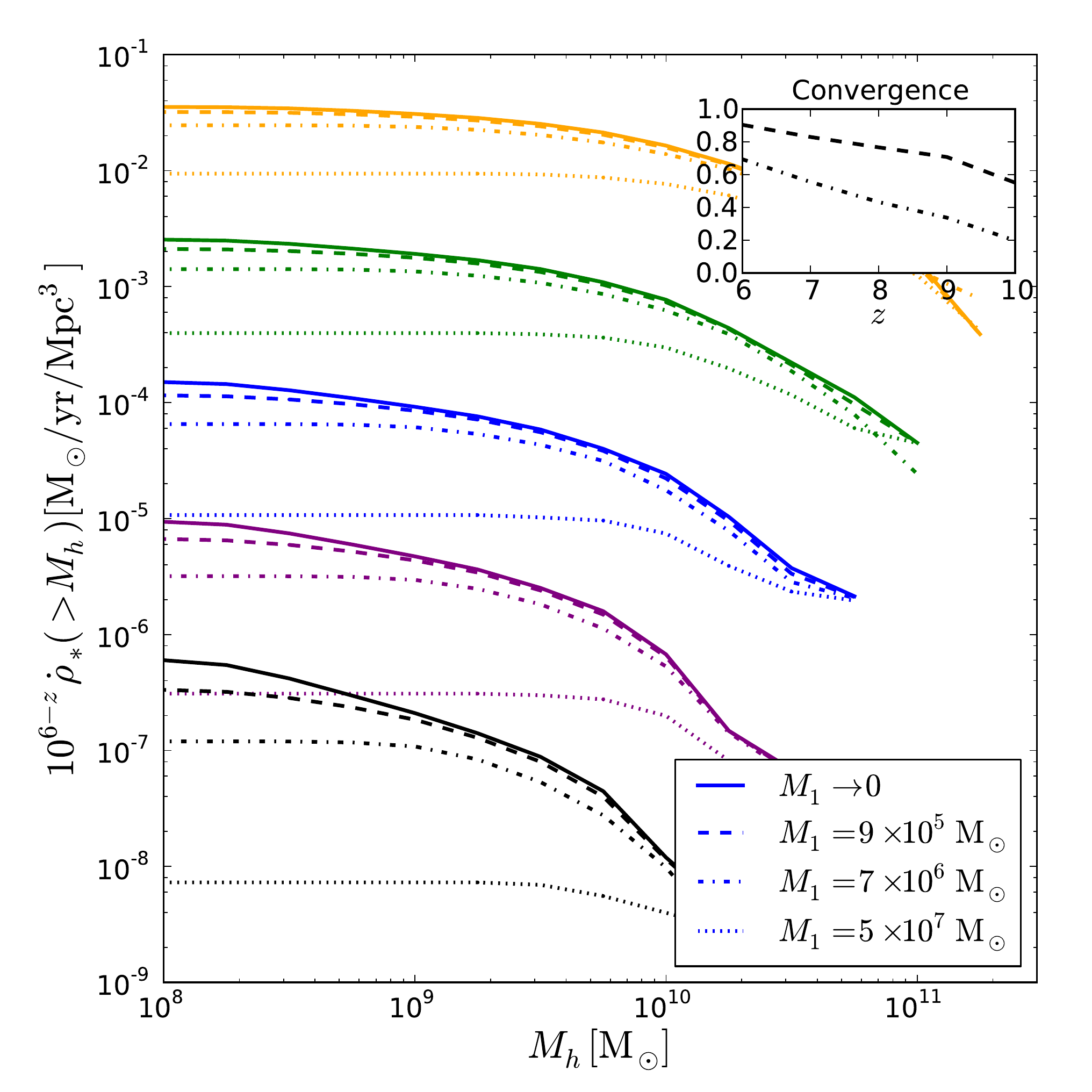}
  \caption{Left: Galaxy luminosity functions for runs with variable mass resolution: \bm\ (dash-dotted lines, total equivalent particle mass $M_1=7\times10^6\Msun$) and B20HR.sf1.uv2.bw10 (dashed lines, total equivalent particle mass $M_1=9\times10^5\Msun$). Right: cumulative star formation rate density in halos vs the lowest halo mass for the same two runs, plus the lowest resolution run B20LR.sf1.uv2.bw10 (dotted lines, total equivalent particle mass $M_1=5\times10^7\Msun$). Solid lines show the extrapolation of these 3 simulations to the limit $M_1\rightarrow0$. The insert plots the degree of convergence in the total star formation rate for \bm\ and B20HR.sf1.uv2.bw10 runs.\label{fig:conv}}
\end{figure}

In order to test mass convergence in our simulations, we compare 3 $20h^{-1}\dim{Mpc}$ runs with different particle counts (and, hence, mass resolutions): the lowest resolution B20LR.sf1.uv2.bw10 ($256^3$ particles, total equivalent particle mass $M_1=5\times10^7\Msun$), the fiducial \bm\ ($512^3$ particles, total equivalent particle mass $M_1=7\times10^6\Msun$), and the highest resolution B20HR.sf1.uv2.bw10 ($1024^3$ particles, total equivalent particle mass $M_1=9\times10^5\Msun$). Figure \ref{fig:conv} shows the luminosity functions at $z=6-10$ (except for B20LR.sf1.uv2.bw10) on the left panel and the cumulative star formation rate density on the right panel.

In numerical analysis it is customary to extrapolate the results at finite resolution in order to estimate the convergence. We extrapolate the cumulative star formation rate density $\dot\rho_*(<M_h|M_1)$ as a function of the simulation resolution $M_1$ to the limit $M_1\rightarrow0$ by adopting the functional form
\begin{equation}
  \dot\rho_*(<M_h|M_1) = \dot\rho_*(<M_h|0)\exp(-\alpha(M_h) M_1^\beta(M_h))
  \label{eq:conv}
\end{equation}
and fitting three parameters $\dot\rho_*(<M_h|0)$, $\alpha(M_h)$, and $\beta(M_h)$ from the three simulation values for $M_1=9\times10^5\Msun$, $7\times10^6\Msun$, and $5\times10^7\Msun$ in each bin of halo mass $M_h$. Admittedly, the adopted functional form is somewhat arbitrary; we checked several different functional forms and found this one to provide the best fits to the simulation results. It can also be justified by noticing that extrapolation is often more precise in logarithmic space, and ansatz (\ref{eq:conv}) is a simple power-law extrapolation in the log space.

With such extrapolation, we can estimate the completeness of our simulations for the total star formation rate as a function of redshift. The highest resolution B20HR.sf1.uv2.bw10 accounts for 90\% of all star formation at $z=6$ and for 60\% at $z=10$. Our fiducial runs include 70\% of all star formation at $z=6$ but only 20\% of all star formation at $z=10$. For the total, integrated over time production of ionizing radiation, the highest resolution run B20HR.sf1.uv2.bw10 misses 20\% of all ionizing photons, while the fiducial runs miss 45\%.

\bibliographystyle{apj}
\bibliography{ng-bibs/self,ng-bibs/ism,ng-bibs/misc,ng-bibs/sims,ng-bibs/sfr,ng-bibs/rei,ng-bibs/dsh,ng-bibs/qlf,ng-bibs/gals,ng-bibs/igm,ng-bibs/reisam}

\end{document}